\documentclass[preprintnumbers,amsmath,amssymbm,prd]{revtex4}
\usepackage{epsfig}
\usepackage{graphicx}
\usepackage{amssymb}

\begin{document}
\title{Analytic treatment of the system of a Kerr-Newman black hole and a charged massive scalar
field}
%\title{Analytic treatment of the composed Kerr-Newman-black-hole-charged-massive-scalar-field bomb}
%%in the eikonal large-mass regime
\author{Shahar Hod}
\affiliation{The Ruppin Academic Center, Emeq Hefer 40250, Israel}
\affiliation{ }
\affiliation{The Hadassah Institute, Jerusalem
91010, Israel}
\date{\today}

\begin{abstract}
\ \ \ Charged rotating Kerr-Newman black holes are known to be
superradiantly unstable to perturbations of charged massive bosonic
fields whose proper frequencies lie in the bounded regime
$0<\omega<\text{min}\{\omega_{\text{c}}\equiv
m\Omega_{\text{H}}+q\Phi_{\text{H}},\mu\}$ [here
$\{\Omega_{\text{H}},\Phi_{\text{H}}\}$ are respectively the angular
velocity and electric potential of the Kerr-Newman black hole, and
$\{m,q,\mu\}$ are respectively the azimuthal harmonic index, the
charge coupling constant, and the proper mass of the field]. In this
paper we study analytically the complex resonance spectrum which
characterizes the dynamics of linearized charged massive scalar
fields in a near-extremal Kerr-Newman black-hole spacetime.
Interestingly, it is shown that near the critical frequency
$\omega_{\text{c}}$ for superradiant amplification and in the
eikonal large-mass regime, the superradiant instability growth rates
of the explosive scalar fields are characterized by a non-trivial
(non-monotonic) dependence on the dimensionless charge-to-mass ratio
$q/\mu$. In particular, for given parameters $\{M,Q, J\}$ of the
central Kerr-Newman black hole, we determine analytically the
optimal charge-to-mass ratio $q/\mu$ of the explosive scalar field
which {\it maximizes} the growth rate of the superradiant
instabilities in the composed
Kerr-Newman-black-hole-charged-massive-scalar-field system.
\end{abstract}
\bigskip
\maketitle

%]

\section{Introduction}

Recent analytical \cite{Hodkn} and numerical \cite{Herkn} studies of
the coupled Einstein-Maxwell-Klein-Gordon field equations have
revealed that, thanks to the intriguing mechanism of superradiance
in curved black-hole spacetimes \cite{Zel,PressTeu2,Bekc}, charged
rotating black holes can support stationary bound-state
configurations of charged massive bosonic (integer-spin) fields
which are everywhere regular outside the black-hole horizon
\cite{Notekr,Hodkr,Herkr}.

These stationary bosonic field configurations \cite{Hodkn,Herkn} are
characterized by proper frequencies which coincide with the critical
(threshold) frequency $\omega_{\text{c}}$ for the superradiant
scattering phenomenon in the black-hole spacetime
\cite{Zel,PressTeu2,Bekc}. In particular, stationary charged field
configurations linearly coupled to a charged rotating Kerr-Newman
black hole of mass $M$, electric charge $Q$, and angular momentum
$J=Ma$, are characterized by the simple relation
\cite{Hodkn,Herkn,Noteun}
\begin{equation}\label{Eq1}
\omega_{\text{field}}=\omega_{\text{c}}\equiv
m\Omega_{\text{H}}+q\Phi_{\text{H}}\  ,
\end{equation}
where $\{\omega_{\text{field}},m,q\}$ are respectively the proper
frequency, the azimuthal harmonic index, and the charge coupling
constant of the stationary charged scalar field mode
\cite{Notedim1}, and \cite{Chan}
\begin{equation}\label{Eq2}
\Omega_{\text{H}}={{a}\over{r^2_++a^2}}\ \ \ \ ; \ \ \ \
\Phi_{\text{H}}={{Qr_+}\over{r^2_++a^2}}\
\end{equation}
are respectively the angular velocity and electric potential of the
Kerr-Newman black hole.

The proper frequencies of these stationary bosonic field
configurations are also characterized by the inequality
\cite{Hodkn,Herkn,Notekr,Hodkr,Herkr}
\begin{equation}\label{Eq3}
\omega^2_{\text{field}}<\mu^2
\end{equation}
(here $\mu$ is the proper mass of the bosonic field
\cite{Notedim2}), a property which guarantees that these external
bound-state configurations cannot radiate their energies to spatial
infinity \cite{Hodkn,Herkn,Notekr,Hodkr,Herkr}.

Interestingly, the stationary bosonic-field configurations
(\ref{Eq1}) studied in \cite{Hodkn,Herkn,Notekr,Hodkr,Herkr} mark
the physical boundary between stable and unstable composed
black-hole-field configurations. In particular, the amplitude of an
external bound-state bosonic field configuration whose proper
frequency is characterized by the inequality
$\omega_{\text{field}}>\omega_{\text{c}}$ is known to decay in time
\cite{PressTeu2,Notemas}, whereas the amplitude of an external
bound-state bosonic field configuration whose proper frequency is
characterized by the property [see Eqs. (\ref{Eq1}) and (\ref{Eq3})]
\begin{equation}\label{Eq4}
0<\omega_{\text{field}}<\text{min}\{\omega_{\text{c}},\mu\}
\end{equation}
is known to grow exponentially over time
\cite{Notemas,Notemir,CarDias}.

The superradiant instability properties of the composed
Kerr-Newman-black-hole-charged-massive-scalar-field system were
studied in the interesting work of Furuhashi and Nambu,\cite{FN}. In
particular, it was found that, in the small frequency $M\omega\ll1$
and small charge-coupling $qQ\ll1$ regime, the growth rate
\cite{Notegwr} of the superradiant instabilities is given by the
simple expression \cite{FN,Notecon}
\begin{equation}\label{Eq5}
\Im\omega={{\mu^4}\over{24}}a(M^2-Q^2)(M\mu-qQ)^5\ \ \ \ \text{for}\
\ \ \ \{M\omega\ll1, M\mu\ll1, qQ\ll1\}\  .
\end{equation}
Inspecting the relation (\ref{Eq5}) for the imaginary part of the
resonant frequency which characterizes the composed
black-hole-charged-field system, one realizes that, in the small
frequency $M\omega\ll1$ regime, the characteristic growth rate of
the superradiant instabilities is a monotonically {\it decreasing}
function of the dimensionless quantity $qQ$. That is, for given
values $\{M,Q,a\}$ of the black-hole physical parameters,
$\Im\omega$ is found to be a monotonically decreasing function of
the charge coupling parameter $q$ which characterizes the explosive
scalar fields.

The main goal of the present paper is to analyze the instability
properties of the composed
Kerr-Newman-black-hole-charged-massive-scalar-field system in the
regime of large field frequencies. To this end, we shall study the
complex resonance spectrum which characterizes the dynamics of the
charged massive scalar fields in the near-extremal charged spinning
Kerr-Newman black-hole spacetime. In particular, below we shall
determine analytically the characteristic growth rates of the
superradiant instabilities near the threshold (critical) frequency
$\omega_{\text{c}}$ [see Eq. (\ref{Eq1})] \cite{Notethrs}.
Interestingly, as we shall explicitly show in the present analysis,
the superradiant instability growth rates of the explosive charged
massive scalar fields near the critical frequency (\ref{Eq1}) are
characterized by a non-trivial ({\it non}-monotonic) dependence on
the dimensionless black-hole-field charge coupling parameter $qQ$.
In particular, for given parameters $\{M,Q, a\}$ of the central
Kerr-Newman black hole, we shall determine analytically the optimal
charge-to-mass ratio $q/\mu$ of the explosive scalar field which
{\it maximizes} the growth rate of the superradiant instabilities in
this composed Kerr-Newman-black-hole-charged-massive-scalar-field
system.

\section{Description of the system}

We shall study analytically the superradiant instability properties
of a physical system which is composed of a charged massive scalar
field $\Psi$ which is linearly coupled to a charged spinning
near-extremal Kerr-Newman black hole. In terms of the familiar
Boyer-Lindquist coordinates $(t,r,\theta,\phi)$, the line element
which describes the external spacetime of a Kerr-Newman black hole
of mass $M$, electric charge $Q$, and angular momentum per unit mass
$a=J/M$ is given by \cite{Chan}
\begin{eqnarray}\label{Eq6}
ds^2=-{{\Delta}\over{\rho^2}}(dt-a\sin^2\theta
d\phi)^2+{{\rho^2}\over{\Delta}}dr^2+\rho^2
d\theta^2+{{\sin^2\theta}\over{\rho^2}}\big[a
dt-(r^2+a^2)d\phi\big]^2\  ,
\end{eqnarray}
where $\Delta\equiv r^2-2Mr+a^2+Q^2$ and $\rho^2\equiv
r^2+a^2\cos^2\theta$. The zeroes of the metric function $\Delta$,
\begin{equation}\label{Eq7}
r_{\pm}=M\pm(M^2-a^2-Q^2)^{1/2}\  ,
\end{equation}
determine the radii of the black-hole (event and inner) horizons.

The dynamics of a linearized scalar field of mass $\mu$ and charge
coupling constant $q$ in the Kerr-Newman black-hole spacetime is
governed by the familiar Klein-Gordon wave equation \cite{Teuk,Stro}
\begin{equation}\label{Eq8}
[(\nabla^\nu-iqA^\nu)(\nabla_{\nu}-iqA_{\nu}) -\mu^2]\Psi=0\  ,
\end{equation}
where $A_{\nu}$ is the electromagnetic potential of the charged
black hole. It is convenient to decompose the scalar field
eigenfunction $\Psi(t,r,\theta,\phi)$ in the form
\cite{Teuk,Stro,Notedec}
\begin{equation}\label{Eq9}
\Psi=\sum_{l,m}e^{im\phi}{S_{lm}}(\theta;a\sqrt{\mu^2-\omega^2}){R_{lm}}(r;M,Q,a,\mu,q,\omega)e^{-i\omega
t}\ ,
\end{equation}
where $R_{lm}$ is the radial part of the scalar eigenfunction and
$S_{lm}$ is the angular part of the scalar eigenfunction.
Substituting the scalar field decomposition (\ref{Eq9}) back into
the Klein-Gordon wave equation (\ref{Eq8}) and using the line
element (\ref{Eq6}) of the curved Kerr-Newman black-hole spacetime,
one obtains \cite{Teuk,Stro} two coupled ordinary differential
equations [see Eqs. (\ref{Eq10}) and (\ref{Eq12}) below] of the
confluent Heun type \cite{Heun,Fiz1,Teuk,Abram,Stro,Hodasy} for the
angular and radial parts of the charged massive scalar
eigenfunction.

The angular (spheroidal harmonic) functions $S_{lm}(\theta)$ satisfy
the ordinary differential equation
\cite{Heun,Fiz1,Teuk,Abram,Stro,Hodasy}
\begin{eqnarray}\label{Eq10}
{1\over {\sin\theta}}{{d}\over{\theta}}\Big(\sin\theta {{d
S_{lm}}\over{d\theta}}\Big) +\Big[K_{lm}+a^2(\mu^2-\omega^2)
-a^2(\mu^2-\omega^2)\cos^2\theta-{{m^2}\over{\sin^2\theta}}\Big]S_{lm}=0\
.
\end{eqnarray}
This differential equation determines the discrete family of angular
eigenvalues $\{K_{lm}\}$ which characterize the regular
\cite{Noteran} angular eigenfunctions $\{S_{lm}(\theta)\}$
\cite{Heun,Fiz1,Teuk,Abram,Stro,Hodasy}. For later purposes we note
that, in the asymptotic $m\gg1$ regime, the angular eigenvalues of
the spheroidal differential equation (\ref{Eq10}) are characterized
by the simple asymptotic behavior \cite{Hodpp,Notewlm}
\begin{equation}\label{Eq11}
K_{mm}=m^2[1+O(m^{-1})]-a^2(\mu^2-\omega^2)\  .
\end{equation}

The radial part of the Klein-Gordon wave equation (\ref{Eq8}) in the
Kerr-Newman black-hole spacetime is given by
\cite{Teuk,Stro,Notecoup}
\begin{equation}\label{Eq12}
{{d}
\over{dr}}\Big(\Delta{{dR_{lm}}\over{dr}}\Big)+\Big[{{H^2}\over{\Delta}}
+2ma\omega-\mu^2(r^2+a^2)-K_{lm}\Big]R_{lm}=0\ ,
\end{equation}
where
\begin{equation}\label{Eq13}
H\equiv \omega(r^2+a^2)-ma-qQr\  .
\end{equation}
The differential equation (\ref{Eq12}), which determines the radial
behavior of the charged massive scalar fields in the charged
spinning Kerr-Newman black-hole spacetime, is supplemented by the
physically motivated boundary condition of purely ingoing scalar
waves (as measured by a comoving observer) at the outer horizon of
the Kerr-Newman black hole \cite{Hodkn,Herkn,Notemas,Notebr}:
\begin{equation}\label{Eq14}
R(r\to r_+)\sim e^{-i(\omega-\omega_{c})y}\  ,
\end{equation}
where the ``tortoise" radial coordinate $y$ is defined by the
relation $dy/dr=(r^2+a^2)/\Delta$ \cite{Noteho}. In addition,
bound-state configurations of the charged massive scalar fields in
the Kerr-Newman black-hole spacetime are characterized by radial
eigenfunctions which, in the small frequency $\omega^2<\mu^2$ regime
[see Eq. (\ref{Eq3})], decay exponentially fast at spatial infinity
\cite{Hodkn,Herkn,Notemas}:
\begin{equation}\label{Eq15}
R(r\to\infty)\sim {{1}\over{r}}e^{-\sqrt{\mu^2-\omega^2}r}\  .
\end{equation}

The radial differential equation (\ref{Eq12}), supplemented by the
boundary conditions (\ref{Eq14}) and (\ref{Eq15}), single out a
discrete spectrum of complex resonant frequencies
$\{\omega(\mu,q,l,m,M,Q,a;n)\}$ \cite{Notenr} which characterize the
dynamics of the charged massive scalar fields in the charged
rotating Kerr-Newman black-hole spacetime. In particular, resonant
frequencies whose imaginary parts are positive are associated with
the exponentially growing superradiant instabilities
\cite{Notemas,Notemir,CarDias} which characterize the composed
black-hole-scalar-field system [see Eq. (\ref{Eq9})]. As we shall
show below, for near-extremal Kerr-Newman black holes in the regime
$(r_+-r_-)/r_+\ll1$, the characteristic complex resonance spectrum
of the composed Kerr-Newman-charged-massive-scalar-field system can
be studied analytically in the vicinity of the critical resonant
frequency $\omega_{\text{c}}$ [see Eq. (\ref{Eq1})] \cite{Notethrs}.

\section{The resonance equation and its regime of validity}

In the present section we shall study the differential equation
(\ref{Eq12}) which determines the spatial behavior of the radial
scalar eigenfunctions. In particular, we shall derive a resonance
condition [see Eq. (\ref{Eq42}) below] for the complex
eigenfrequencies which characterize the dynamics of the charged
massive scalar fields in the spacetime of a near-extremal charged
rotating Kerr-Newman black hole.

The resonance equation for the complex resonant frequencies which
characterize the dynamics of {\it neutral} scalar fields in the
spacetime of a {\it neutral} near-extremal Kerr black hole was
derived in \cite{HodHod}. It is important to emphasize that the
analysis presented in \cite{HodHod} is restricted to the regime
$M\mu=O(1)$ of moderate field masses. In the present study we shall
generalize the analysis of \cite{HodHod} to the regime of {\it
charged} massive scalar fields propagating in the spacetime of a
{\it charged} near-extremal Kerr-Newman black hole. In addition,
below we shall extend the analysis of \cite{HodHod} to the regime
$M\mu\gg1$ of large field masses \cite{Notecare}.

It is convenient to express the physical quantities which
characterize the composed
Kerr-Newman-black-hole-linearized-charged-massive-scalar-field
system in terms of the dimensionless variables \cite{Teuk,Stro}
\begin{equation}\label{Eq16}
x\equiv {{r-r_+}\over {r_+}}\ \ \ ;\ \ \
\tau\equiv{{r_+-r_-}\over{r_+}}\ \ \ ;\ \ \ k\equiv 2\omega r_+-qQ\
\ \ ; \ \ \ \varpi\equiv{{\omega-\omega_{\text{c}}}\over{2\pi
T_{\text{BH}}}}\  ,
\end{equation}
where $T_{\text{BH}}=(r_+-r_-)/4\pi(r^2_++a^2)$ is the
Bekenstein-Hawking temperature of the charged spinning Kerr-Newman
black hole. Substituting (\ref{Eq16}) into (\ref{Eq12}), one finds
the differential equation
\begin{equation}\label{Eq17}
x(x+\tau){{d^2R}\over{dx^2}}+(2x+\tau){{dR}\over{dx}}+UR=0\
\end{equation}
for the radial eigenfunctions of the charged massive scalar fields
in the Kerr-Newman black-hole spacetime, where
\begin{equation}\label{Eq18}
U=U(x;\mu,q,\omega,l,m,M,Q,a)={{[\omega
r_+x^2+kx+\varpi\tau/2]^2}\over{x(x+\tau)}}-K+2ma\omega-\mu^2[r^2_+(1+x)^2+a^2]\
.
\end{equation}

The radial equation (\ref{Eq17}) can be solved analytically in the
two asymptotic regions $x\ll1$ and
$x\gg\text{max}\{\tau,M(\omega_{\text{c}}-\omega)\}$ \cite{HodHod}.
Note, in particular, that in the double asymptotic regime
\cite{Notedar}
\begin{equation}\label{Eq19}
\tau\ll1\ \ \ \ \text{and}\ \ \ \ M(\omega_{\text{c}}-\omega)\ll1\ ,
\end{equation}
one can use a standard matching procedure in the overlapping region
$\text{max}\{\tau,M(\omega_{\text{c}}-\omega)\}\ll x\ll 1$ in order
to determine the complex resonant frequencies
$\{\omega(\mu,q,l,m,M,Q,a;n)\}$ which characterize the dynamics of
the charged massive scalar fields in the charged spinning
Kerr-Newman black-hole spacetime.

We shall first solve the radial differential equation (\ref{Eq17})
in the region
\begin{equation}\label{Eq20}
x\ll 1\  ,
\end{equation}
in which case one can use the near-horizon approximation $U\to
U_{\text{near}}\equiv
(kx+\varpi\tau/2)^2/[x(x+\tau)]-K+2ma\omega-\mu^2(r^2_++a^2)$ for
the effective radial potential in (\ref{Eq17}). The near-horizon
radial solution of (\ref{Eq17}) which respects the physically
motivated boundary condition (\ref{Eq14}) at the outer horizon of
the Kerr-Newman black hole can be expressed in terms of the
hypergeometric function \cite{Abram,HodHod,Morse}:
\begin{equation}\label{Eq21}
R(x)=x^{-i{{\varpi}\over{2}}}\Big({x\over
\tau}+1\Big)^{i{{\varpi}\over{2}}-ik}{_2F_1}({1\over
2}+i\delta-ik,{1\over 2}-i\delta-ik;1-i\varpi;-x/\tau)\  ,
\end{equation}
where
\begin{equation}\label{Eq22}
\delta^2\equiv -K-{1\over 4}+2ma\omega+k^2-\mu^2(r^2_++a^2)\ .
\end{equation}
It proves useful to write the near-horizon radial solution
(\ref{Eq21}) in the form (see Eq. 15.3.7 of \cite{Abram})
\begin{eqnarray}\label{Eq23}
R(x)&=&x^{-i{{\varpi}\over{2}}}\Big({x\over
\tau}+1\Big)^{i{{\varpi}\over{2}}-ik}\Big[
{{\Gamma(1-i\varpi)\Gamma(2i\delta)}\over{\Gamma({1/2}+i\delta-ik)\Gamma({1/2}+i\delta+ik-i\varpi)}}
\Big({{x}\over{\tau}}\Big)^{-1/2+i\delta+ik} \nonumber \\&& \times
{_2F_1}({1\over 2}-i\delta-ik,{1\over
2}-i\delta-ik+i\varpi;1-2i\delta;-\tau/x)+(\delta\to -\delta)\Big]\
,
\end{eqnarray}
where the notation $(\delta\to -\delta)$ means ``replace $\delta$ by
$-\delta$ in the preceding term". Using the simple asymptotic
behavior (see Eq. 15.1.1 of \cite{Abram})
\begin{equation}\label{Eq24}
_2F_1(a,b;c;z)\to 1\ \ \ \text{for}\ \ \ {{ab}\over{c}}\cdot z\to 0\
\end{equation}
of the hypergeometric function, one finds from (\ref{Eq23}) the
expression
\begin{eqnarray}\label{Eq25}
R(x)={{\Gamma(1-i\varpi)\Gamma(2i\delta)\tau^{1/2-i\delta-i\varpi/2}}\over{\Gamma({1/2}+i\delta-ik)
\Gamma({1/2}+i\delta+ik-i\varpi)}}x^{-{{1}\over{2}}+i\delta}
+(\delta\to -\delta)\
\end{eqnarray}
for the radial eigenfunction of the charged massive scalar fields in
the intermediate region
\begin{equation}\label{Eq26}
\tau\times \text{max}(m,\varpi)\ll x\ll 1\  .
\end{equation}

We shall next solve the radial differential equation (\ref{Eq17}) in
the region
\begin{equation}\label{Eq27}
x\gg \text{max}(\tau,\varpi\tau/m)\  ,
\end{equation}
in which case one can replace (\ref{Eq17}) by
\begin{equation}\label{Eq28}
x^2{{d^2R}\over{dx^2}}+2x{{dR}\over{dx}}+U_{\text{far}}R=0\  ,
\end{equation}
where the effective potential in (\ref{Eq28}) is given by $U\to
U_{\text{{far}}}=(\omega
r_+x+k)^2-K+2ma\omega-\mu^2[r^2_+(1+x)^2+a^2]$. The radial solution
of (\ref{Eq28}) can be expressed in terms of the confluent
hypergeometric function \cite{Abram,HodHod,Morse}:
\begin{equation}\label{Eq29}
R(x)=N_1\times(2\epsilon)^{{1\over 2}+i\delta}x^{-{1\over
2}+i\delta}e^{-\epsilon x}{_1F_1}({1\over
2}+i\delta-\kappa,1+2i\delta,2\epsilon x)+N_2\times(\delta\to
-\delta)\ ,
\end{equation}
where we have used here the dimensionless variables
\begin{equation}\label{Eq30}
\epsilon\equiv \sqrt{\mu^2-\omega^2}r_+\ \ \ \ ; \ \ \ \
\kappa\equiv {{\omega kr_+-(\mu r_+)^2}\over{\epsilon}}\ .
\end{equation}
As we shall show below, the normalization constants $\{N_1,N_2\}$ of
the radial solution (\ref{Eq29}) can be determined analytically by a
standard matching procedure. Using the simple asymptotic behavior
(see Eq. 13.1.2 of \cite{Abram})
\begin{equation}\label{Eq31}
_1F_1(a,b,z)\to 1\ \ \ \text{for}\ \ \ {{a}\over{b}}\cdot z\to 0\
\end{equation}
of the confluent hypergeometric function, one finds from
(\ref{Eq29}) the expression
\begin{equation}\label{Eq32}
R(x)=N_1\times(2\epsilon)^{{1\over 2}+i\delta}x^{-{1\over
2}+i\delta}+N_2\times(\delta\to -\delta)\
\end{equation}
for the radial eigenfunction of the charged massive scalar fields in
the intermediate region
\begin{equation}\label{Eq33}
\tau\times\text{max}(1,\varpi/m)\ll x\ll m^{-1}\  .
\end{equation}

From Eqs. (\ref{Eq26}) and (\ref{Eq33}) one learns that, for
near-extremal charged spinning Kerr-Newman black holes in the regime
$\tau\ll1$, there is an overlap radial region which is determined by
the strong inequalities
\begin{equation}\label{Eq34}
\tau\times\text{max}(m,\varpi)\ll x_o\ll m^{-1}\  ,
\end{equation}
in which the expressions (\ref{Eq21}) and (\ref{Eq29}) for the
radial scalar eigenfunction $R(x)$ are both valid. Note, in
particular, that the two expressions (\ref{Eq25}) and (\ref{Eq32})
for the radial eigenfunction in the overlap region (\ref{Eq34}) have
the same functional dependence on the dimensionless radial
coordinate $x$. Thus, one can determine the normalization constants
$N_1$ and $N_2$ of the radial eigenfunction (\ref{Eq29}) by matching
the expressions (\ref{Eq25}) and (\ref{Eq32}) in their overlap
region (\ref{Eq34}). This matching procedure yields
\begin{equation}\label{Eq35}
N_1(\delta)={{\Gamma(1-i\varpi)\Gamma(2i\delta)}\over{\Gamma({1\over
2}+i\delta-ik)\Gamma({1\over 2}+i\delta+ik-i\varpi)}}\tau^{{1\over
2}-i\delta-i{{\varpi}\over{2}}}(2\epsilon)^{-{1\over 2}-i\delta}\ \
\ \ {\text{and}} \ \ \ \ N_2(\delta)=N_1(-\delta)\  .
\end{equation}

We shall now derive the characteristic equation which determines the
complex resonant frequencies of the composed
Kerr-Newman-black-hole-charged-massive-scalar-field system. We first
point out that the radial eigenfunction (\ref{Eq29}) of the charged
massive scalar fields is characterized by the asymptotic behavior
(see Eq. 13.5.1 of \cite{Abram})
\begin{eqnarray}\label{Eq36}
R(x\to\infty)&\to&
\Big[N_1\times(2\epsilon)^{\kappa}{{\Gamma(1+2i\delta)}\over{\Gamma({1\over
2}+i\delta+\kappa)}}x^{-1+\kappa}(-1)^{-{1\over
2}-i\delta+\kappa}+N_2\times(\delta\to -\delta)\Big]e^{-\epsilon x}
\nonumber \\&& +
\Big[N_1\times(2\epsilon)^{-\kappa}{{\Gamma(1+2i\delta)}\over{\Gamma({1\over
2}+i\delta-\kappa)}}x^{-1-\kappa}+N_2\times(\delta\to
-\delta)\Big]e^{\epsilon x}\
\end{eqnarray}
at spatial infinity. Taking cognizance of the boundary condition
(\ref{Eq15}), which characterizes the spatial behavior of the
bound-state radial scalar eigenfunctions at asymptotic infinity, one
realizes that the coefficient of the exploding exponent $e^{\epsilon
x}$ in the asymptotic expression (\ref{Eq36}) must vanish:
\begin{eqnarray}\label{Eq37}
N_1\times(2\epsilon)^{-\kappa}{{\Gamma(1+2i\delta)}\over{\Gamma({1\over
2}+i\delta-\kappa)}}x^{-1-\kappa}+N_2\times(\delta\to -\delta)=0\  .
\end{eqnarray}
Substituting into (\ref{Eq37}) the normalization constants $N_1$ and
$N_2$ [see Eq. (\ref{Eq35})], one finds the resonance equation
\begin{equation}\label{Eq38}
\Big[{{\Gamma(-2i\delta)}\over{\Gamma(2i\delta)}}\Big]^2{{\Gamma({1\over
2}+i\delta-ik)\Gamma({1\over 2}+i\delta-\kappa)\Gamma({1\over
2}+i\delta+ik-i\varpi)}\over{\Gamma({1\over
2}-i\delta-ik)\Gamma({1\over 2}-i\delta-\kappa)\Gamma({1\over
2}-i\delta+ik-i\varpi)}}\big(2\epsilon\tau\big)^{2i\delta}=1\
\end{equation}
which determines the complex resonant frequencies of the charged
massive scalar fields in the near-extremal charged rotating
Kerr-Newman black-hole spacetime.

We note that the resonance equation (\ref{Eq38}) can be simplified
in the regime
\begin{equation}\label{Eq39}
\tau\ll {{\bar\omega}\over{m}}
\end{equation}
of near-extremal Kerr-Newman black holes, where here \cite{Notenkn}
\begin{equation}\label{Eq40}
{\bar\omega}\equiv
{{(r^2_++a^2)(\omega-\omega_{\text{c}})}\over{r_+}}\
\end{equation}
is a dimensionless parameter which quantifies the distance between
the proper frequency of the charged massive scalar field and the
critical frequency (\ref{Eq1}) \cite{Notethrs} for superradiant
scattering in the charged rotating Kerr-Newman black-hole spacetime.
In particular, in the near-extremal regime (\ref{Eq39}), one can use
the approximated relation \cite{Abram,Notenkn}
\begin{equation}\label{Eq41}
{{\Gamma({1\over 2}+i\delta+ik-i\varpi)}\over{\Gamma({1\over
2}-i\delta+ik-i\varpi)}}=(-i\varpi)^{2i\delta}[1+O(m/\varpi)]\
\end{equation}
for the Gamma functions that appear in the resonance equation
(\ref{Eq38}). Substituting (\ref{Eq41}) into (\ref{Eq38}), one finds
the resonance condition
\begin{equation}\label{Eq42}
\Big[{{\Gamma(-2i\delta)}\over{\Gamma(2i\delta)}}\Big]^2{{\Gamma({1\over
2}+i\delta-ik)\Gamma({1\over 2}+i\delta-\kappa)}\over{\Gamma({1\over
2}-i\delta-ik)\Gamma({1\over
2}-i\delta-\kappa)}}\big(-4i\epsilon{\bar\omega}\big)^{2i\delta}=1\
.
\end{equation}
It is worth emphasizing again that the resonance equation
(\ref{Eq42}) is valid in the regime [see Eqs. (\ref{Eq16}),
(\ref{Eq34}), (\ref{Eq39}), and (\ref{Eq40})]
\begin{equation}\label{Eq43}
m\tau\ll{\bar\omega}\ll m^{-1}\  .
\end{equation}

In the next section we shall show that, for $\delta\in\mathbb{R}$
\cite{Notedel}, the (rather cumbersome) resonance equation
(\ref{Eq42}) yields a remarkably simple expression for the
dimensionless ratio
$\omega_{\text{I}}/(\omega_{\text{R}}-\omega_{\text{c}})$, where
$\{\omega_{\text{R}},\omega_{\text{I}}\}$ are respectively the real
and imaginary parts of the complex resonant frequencies which
characterize the dynamics of the charged massive scalar fields in
the near-extremal charged spinning Kerr-Newman black-hole spacetime.

\section{The superradiant instability spectrum of the composed Kerr-Newman-black-hole-charged-massive-scalar-field system}

Taking cognizance of the derived resonance equation (\ref{Eq42}),
one finds that the resonant frequencies of the composed
Kerr-Newman-black-hole-charged-massive-scalar-field system in the
regime (\ref{Eq43}) can be expressed in the compact form
\begin{equation}\label{Eq44}
\bar\omega={\cal R}\times{\cal J}\  ,
\end{equation}
where \cite{NoteGam,Noteintg}
\begin{equation}\label{Eq45}
{\cal R}\equiv {{e^{-\pi n/\delta}}\over{4\epsilon}}
\Big\{\Big[{{\Gamma(2i\delta)}\over{\Gamma(-2i\delta)}}\Big]^2{{\Gamma({1\over
2}-i\delta-\kappa)}\over{\Gamma({1\over
2}+i\delta-\kappa)}}\Big\}^{1/2i\delta}\in\mathbb{R}\
\end{equation}
and
\begin{equation}\label{Eq46}
{\cal J}\equiv i\Big[{{\Gamma({1\over
2}-i\delta-ik)}\over{\Gamma({1\over
2}+i\delta-ik)}}\Big]^{1/2i\delta}\in\mathbb{C}\  .
\end{equation}
Equations (\ref{Eq44}), (\ref{Eq45}), and (\ref{Eq46}) imply the
relations
\begin{equation}\label{Eq47}
\bar\omega_{\text{I}}={\cal R}\times{\cal J}_{\text{I}}\ \ \ \
\text{and}\ \ \ \ \bar\omega_{\text{R}}={\cal R}\times{\cal
J}_{\text{R}}\  ,
\end{equation}
which, in turn, yield the remarkably simple dimensionless ratio
\begin{equation}\label{Eq48}
%{\cal F}(l,m,\alpha)\equiv
{{\omega_{\text{I}}}\over{\omega_{\text{R}}-\omega_{\text{c}}}}
%{{\bar\omega_{\text{I}}}\over{\bar\omega_{\text{R}}}}
={{{\cal J}_{\text{I}}}\over{{\cal J}_{\text{R}}}}\
\end{equation}
for the resonant frequencies of the charged massive scalar fields in
the near-extremal Kerr-Newman black-hole spacetime.

In the next section we shall study the eikonal large-mass $M\mu\gg1$
regime \cite{Notemmu} of the composed
Kerr-Newman-black-hole-charged-massive-scalar-field system. In
particular, below we shall show that the characteristic
dimensionless ratio
${{\omega_{\text{I}}}/({\omega_{\text{R}}-\omega_{\text{c}}}})$ [see
Eq. (\ref{Eq48})] can be expressed in a remarkably compact form in
this large-mass regime.

\section{The eikonal large-mass $M\mu\gg1$ regime}

In the present section we shall analyze the asymptotic large-mass
regime
\begin{equation}\label{Eq49}
M\mu\gg1
\end{equation}
of the composed Kerr-Newman-black-hole-charged-massive-scalar-field
system. In the asymptotic regime (\ref{Eq49}), one can use the
approximated relation \cite{Abram,Notemmu}
\begin{equation}\label{Eq50}
{{\Gamma({1\over 2}-i\delta-ik)}\over{\Gamma({1\over
2}+i\delta-ik)}}=e^{(2i-\pi)\delta}(k+\delta)^{-i(k+\delta)}(k-\delta)^{i(k-\delta)}[1+e^{-2\pi(k-\delta)}]
[1+O(m^{-1})]\
\end{equation}
for the Gamma functions that appear in the expression (\ref{Eq46})
for ${\cal J}$. Substituting (\ref{Eq50}) into (\ref{Eq46}), one
finds
\begin{equation}\label{Eq51}
{\cal
J}=-e(k+\delta)^{-(k+\delta)/2\delta}(k-\delta)^{(k-\delta)/2\delta}[1+e^{-2\pi(k-\delta)}]^{1/2i\delta}\
,
\end{equation}
which yields the remarkably simple dimensionless relation [see Eq.
(\ref{Eq48})] \cite{Notese}
\begin{equation}\label{Eq52}
%{\cal F}(M\mu\gg1)
{{\omega_{\text{I}}}\over{\omega_{\text{c}}-\omega_{\text{R}}}}={{e^{-2\pi(k-\delta)}}\over{2\delta}}\
\end{equation}
for the characteristic resonant frequencies of the composed
Kerr-Newman-black-hole-charged-massive-scalar-field system in the
eikonal large-mass regime (\ref{Eq49}).

As a consistency check, we shall now compare our large-mass result
(\ref{Eq52}) for the resonant frequencies of the composed
black-hole-field system with the corresponding large-mass result of
Zouros and Eardley \cite{ZouEar}. In their highly important work,
Zouros and Eardley \cite{ZouEar} have performed a WKB analysis for
the specific case of {\it neutral} scalar fields linearly coupled to
a {\it neutral} spinning Kerr black hole in the large-mass
$M\mu\gg1$ regime. In particular, for the case of near-extremal Kerr
black holes in the regime \cite{Notempl}
\begin{equation}\label{Eq53}
a\simeq M\ \ \ \ ; \ \ \ \ l=m\gg1\ \ \ \ ; \ \ \ \  \mu\simeq
\omega\simeq m\Omega_{\text{H}}\simeq m/2M\gg 1\  ,
\end{equation}
Zouros and Eardley \cite{ZouEar} have derived the well known WKB
result \cite{Noteer}
\begin{equation}\label{Eq54}
M\omega_{\text{I}}\propto e^{-2\pi(2-\sqrt{2})M\mu}\  .
\end{equation}
Note that, for near-extremal Kerr black holes, the specific case
(\ref{Eq53}) corresponds to [see Eqs. (\ref{Eq11}), (\ref{Eq16}),
and (\ref{Eq22})]
\begin{equation}\label{Eq55}
k=m\ \ \ \ \text{and}\ \ \ \ \delta={{m}\over{\sqrt{2}}}+O(1)\  .
\end{equation}
Substituting (\ref{Eq55}) into our analytically derived expression
(\ref{Eq52}), one finds the dimensionless ratio
\begin{equation}\label{Eq56}
%{\cal F}(l=m\gg1,\alpha\to m/2\gg1)
{{\omega_{\text{I}}}\over{\omega_{\text{c}}-\omega_{\text{R}}}}={{e^{-2\pi(2-\sqrt{2})M\mu}}
\over{2\sqrt{2}M\mu}}\  ,
\end{equation}
a result which is consistent with the important result (\ref{Eq54})
of Zouros and Eardley \cite{ZouEar} for the specific case of {\it
neutral} scalar fields linearly coupled to a {\it neutral}
near-extremal spinning Kerr black hole.

\section{The optimal charge-to-mass ratio of the explosive scalar fields}

In the present section we shall analyze the functional dependence of
the superradiant instability growth rate (\ref{Eq52}) on the
dimensionless ratio $q/\mu$ which characterizes the explosive
charged massive scalar fields. Taking cognizance of Eqs.
(\ref{Eq1}), (\ref{Eq11}) \cite{Noteklem}, (\ref{Eq16}), and
(\ref{Eq22}), one finds the expression
%\begin{equation}\label{Eq57}
%\delta-k=\sqrt{-K-1/4+2ma\omega_{\text{c}}+(2\omega_{\text{c}}r_+-qQ)^2-\mu^2(r^2_++a^2)}-(2\omega_{\text{c}}r_+-qQ)
%\end{equation}
\begin{equation}\label{Eq57}
\delta-k=\sqrt{(2\omega_{\text{c}}r_+-qQ)^2-(a\omega_{\text{c}}-m)^2-\mu^2r^2_+}-(2\omega_{\text{c}}r_+-qQ)
\end{equation}
for the exponent of (\ref{Eq52}) near the superradiant instability
threshold (\ref{Eq1}) \cite{Notethrs,Notewn} of the near-extremal
Kerr-Newman black holes.
%Interestingly, from Eq. (\ref{Eq52}) one learns that,
%for a given value of the critical field frequency
%$\omega_{\text{c}}$, the superradiant instability growth rate of the
%charged massive scalar fields (that is, the value of
%$\omega_{\text{I}}$) can be maximized by maximizing the value of the
%exponent $\delta-k$.
From Eq. (\ref{Eq57}) one immediately learns that the exponent
$\delta-k$ is a monotonically decreasing function of the mass
parameter $\mu$. Thus, one can maximize the value of the exponent
(\ref{Eq57}) by minimizing (for a given value of the critical field
frequency $\omega_{\text{c}}$) the proper mass of the explosive
scalar field. In particular, taking cognizance of Eq. (\ref{Eq3})
one realizes that, for a given value of the critical field frequency
$\omega_{\text{c}}$ \cite{Notewn}, the exponent (\ref{Eq57}) can be
maximized by taking
\begin{equation}\label{Eq58}
{{\mu}\over{\omega_{\text{c}}}}\to 1^+\  .
% \ \ \ \text{where}\ \ \ \ \omega_{\text{c}}=m\Omega_{\text{H}}+q\Phi_{\text{H}}\  .
\end{equation}
Substituting Eqs. (\ref{Eq1}), (\ref{Eq2}), and (\ref{Eq58}) into
(\ref{Eq57}), and defining the dimensionless quantities
\begin{equation}\label{Eq59}
\gamma\equiv{{qQ}\over{m}}\ \ \ \ ; \ \ \ \ s\equiv {{a}\over{r_+}}\
,
\end{equation}
one finds
\begin{equation}\label{Eq60}
\delta-k=m\cdot{{\sqrt{-s^2(3-s^2)\gamma^2+4s(1-s^2)\gamma+3s^2-1}-[2s+(1-s^2)\gamma]}\over{1+s^2}}\
\end{equation}
for the maximally allowed value of the exponent (\ref{Eq57}) near
the superradiant instability threshold (\ref{Eq1}) of the
near-extremal Kerr-Newman black holes.

For a given value of the dimensionless black-hole rotation parameter
$s$, the superradiant instability growth rate of the charged massive
scalar fields [that is, the value of $\omega_{\text{I}}$, see Eq.
(\ref{Eq52})] can be maximized by maximizing with respect to
$\gamma$ the expression (\ref{Eq60}) for the exponent $\delta-k$
\cite{NoteqQd}. In particular, a simple differentiation of
(\ref{Eq60}) with respect to the dimensionless variable $\gamma$
yields \cite{Notedm}
\begin{equation}\label{Eq61}
%\text{max}_{\gamma}\{\delta-k\}=m\cdot
%{{\sqrt{1+s^2}-2}\over{s(3-s^2)}}\  ,
\text{max}_{\gamma}\{\delta-k\}=-{{m}\over{s(2+\sqrt{1+s^2})}}\  ,
\end{equation}
where this maximally allowed value of the exponent $\delta-k$ is
obtained for
\begin{equation}\label{Eq62}
%\gamma=\gamma^*(s)={{(2-\sqrt{1+s^2})(1-s^2)}\over{s(3-s^2)}}\ .
\gamma=\gamma^*(s)={{1-s^2}\over{s(2+\sqrt{1+s^2})}}\  .
\end{equation}
It is worth noting that the expression (\ref{Eq61}) for
$\text{max}\{\delta-k\}$ is a monotonically increasing function of
the dimensionless black-hole rotation parameter $s$. In particular,
one finds from (\ref{Eq61}) $\text{max}\{\delta-k\}=m(1/\sqrt{2}-1)$
%for the exponent of (\ref{Eq52})
in the $s\to 1$ limit, in agreement
with the highly important result (\ref{Eq54}) of Zouros and Eardley
\cite{ZouEar} for the specific case of {\it neutral} scalar fields
linearly coupled to a {\it neutral} near-extremal spinning Kerr
black hole.

\section{Summary}

The superradiant instability properties of the composed
Kerr-Newman-black-hole-charged-massive-scalar-field system were
studied {\it analytically}. In particular, we have analyzed the
near-critical \cite{Notethrs,Notewn} complex resonance spectrum
which characterizes the dynamics of linearized charged massive
scalar fields in a near-extremal charged spinning Kerr-Newman
black-hole spacetime.

Interestingly, it was shown that in the eikonal large-mass regime
the superradiant instability growth rates of the explosive charged
massive scalar fields are characterized by a non-trivial ({\it
non}-monotonic) dependence on the dimensionless black-hole-field
charge coupling parameter $qQ$ \cite{Notelwlq}. In particular, for
given parameters $\{M,Q, a\}$ of the central near-extremal
Kerr-Newman black hole, the superradiant instability growth rate is
maximized for [see Eqs. (\ref{Eq1}), (\ref{Eq2}), (\ref{Eq58}),
(\ref{Eq59}), and (\ref{Eq62})]
\begin{equation}\label{Eq63}
(qQ)_{\text{optimal}}=m\cdot {{1-s^2}\over{s(2+\sqrt{1+s^2})}}\ \ \
\ \text{and}\ \ \ \ (M\mu)_{\text{optimal}}=m\cdot
{{s^2+\sqrt{1+s^2}}\over{s(2+\sqrt{1+s^2})\sqrt{1+s^2}}}\  .
\end{equation}
These relations yield the dimensionless compact expression
\begin{equation}\label{Eq64}
%{{q}\over{\mu}}={{(2-\sqrt{1+s^2})(1+s^2)\sqrt{1-s^2}}\over{(2-\sqrt{1+s^2})(1-s^2)+s^2(3-s^2)}}
\Big({{q}\over{\mu}}\Big)_{\text{optimal}}={{\sqrt{1-s^4}}\over{s^2+\sqrt{1+s^2}}}
\end{equation}
for the optimal charge-to-mass ratio of the explosive scalar field
which maximizes the growth rate of the superradiant instabilities.
Finally, taking cognizance of Eqs. (\ref{Eq52}), (\ref{Eq60}), and
(\ref{Eq62}), one finds the large-mass expression
\begin{equation}\label{Eq65}
\text{max}\Big\{{{\omega_{\text{I}}}\over{\omega_{\text{c}}-\omega_{\text{R}}}}\Big\}={{\sqrt{1+s^2}}\over{2s\cdot
m}}\times \exp\Big[-{{2\pi}\over{s(2+\sqrt{1+s^2})}}\cdot m\Big]\
\end{equation}
%\begin{equation}\label{Eq65}
%\text{max}\Big\{{{\omega_{\text{I}}}\over{\omega_{\text{c}}-\omega_{\text{R}}}}\Big\}=
%{{s^2+\sqrt{1+s^2}}\over{2s^2(2+\sqrt{1+s^2})\alpha}}
%\times\exp\Big[-2\pi{{\sqrt{1+s^2}}\over{s^2+\sqrt{1+s^2}}}\cdot\alpha\Big]\
%\ \ ; \ \ \ \alpha\equiv (M\mu)_{\text{optimal}}\gg1
%\end{equation}
for the maximum growth rate of the superradiant instabilities in the
composed Kerr-Newman-black-hole-charged-massive-scalar-field bomb.

\bigskip
\noindent
{\bf ACKNOWLEDGMENTS}
\bigskip

This research is supported by the Carmel Science Foundation. I thank
Yael Oren, Arbel M. Ongo, Ayelet B. Lata, and Alona B. Tea for
stimulating discussions.

%\newpage


\begin{thebibliography}{99}
\bibitem{Notekr} It is worth noting that the physical properties
of these stationary composed black-hole-bosonic-field configurations
were studied extensively in recent years, see \cite{Hodkr,Herkr} and
references therein.
\bibitem{Noteun} We shall use natural units in which $G=c=\hbar=1$.
\bibitem{Notedim1} Note that the charge coupling constant of the scalar field stands for $q/\hbar$. Thus,
this field parameter has the dimensions of $($length$)^{-1}$.
\bibitem{Notedim2} Note that the proper mass of the scalar field stands for $\mu/\hbar$. Thus,
this field parameter has the dimensions of $($length$)^{-1}$.
\bibitem{Notemir} It is worth mentioning that one can also trigger the superradiant
instabilities in the composed black-hole-bosonic-field system by
placing a reflecting mirror around the black hole. The physical role
of this external mirror is to prevent the superradiantly amplified
bosonic fields from escaping to spatial infinity. The physical
properties of this composed black-hole-bosonic-field-mirror bomb
were analyzed in \cite{PressTeu2,CarDias}.
\bibitem{Notegwr} The growth rates of the superradiant instabilities are
determined by the imaginary parts of the resonant frequencies which
characterize the composed black-hole-massive-bosonic-field system.
\bibitem{Notecon} It is worth noting that the expression (\ref{Eq5}) of \cite{FN} for the growth rates of the
superradiant instabilities in the small-frequency ($M\omega\ll1$)
small-mass ($M\mu\ll1$) regime refers to the case of linearized
charged massive scalar fields with $l=m=1$ [here $\{l,m\}$ are the
harmonic indices of the scalar field. As shown in \cite{FN}, the
most unstable (exponentially explosive) scalar mode is characterized
by these values of the harmonic parameters].
\bibitem{Notethrs} It is worth emphasizing that the physical significance of the critical
frequency (\ref{Eq1}) stems from the fact that this resonant field
frequency marks the onset of the exponentially growing superradiant
instabilities in the composed
Kerr-Newman-black-hole-charged-massive-scalar-field system.
\bibitem{Notedec} Here $\omega$ is the conserved frequency of the charged massive
scalar field mode, and $\{l,m\}$ are respectively the spheroidal and
azimuthal harmonic indices (with $l\ge |m|$) of the charged scalar
field mode.
\bibitem{Noteran} The angular eigenfunctions $S_{lm}(\theta)$ are assumed to be regular at the
two angular poles $\theta=0$ and $\theta=\pi$
\cite{Heun,Fiz1,Teuk,Abram,Stro,Hodasy}.
\bibitem{Notewlm} It is worth noting that this asymptotic relation is valid in the regime
$-a^2(\mu^2-\omega^2)<m^2$ \cite{Hodpp}. Taking cognizance of
(\ref{Eq3}), one realizes that this inequality is trivially
satisfied by the bound-state resonances of the massive scalar
fields.
\bibitem{Notecoup} It is worth emphasizing the fact that, for spinning Kerr-Newman black holes, the
ordinary differential equations which determine the angular and
radial behaviors of the charged massive scalar fields [see Eqs.
(\ref{Eq10}) and (\ref{Eq12}), respectively] are coupled by the
angular eigenvalues $\{{K_{lm}}\}$.
\bibitem{Notebr} For brevity, we shall henceforth omit the harmonic indices $\{l,m\}$ of the scalar field mode.
\bibitem{Noteho} Note that the near-horizon limit $r\to r_+$ corresponds to $y\to -\infty$.
\bibitem{Notenr} Here the integer $n$ is the resonance parameter of
the exploding scalar field mode [see Eq. (\ref{Eq45}) below]. This
parameter corresponds to the number of nodes of the radial
eigenfunction.
\bibitem{Notecare} As we shall explicitly show below, the
regime of validity of the final resonance equation [see Eq.
(\ref{Eq42}) below] should be determined with great care in the
double limit of small black-hole temperatures and large field
masses.
\bibitem{Notedar} Note that the double asymptotic limit (\ref{Eq19}) corresponds to
Kerr-Newman black holes in the near-extremal $\tau\ll1$ regime, and
to resonant field frequencies near the critical frequency
$\omega_{\text{c}}$ \cite{Notethrs} for the superradiant
amplification phenomenon in the black-hole spacetime.
\bibitem{Notenkn} Note that the strong inequality (\ref{Eq39}) corresponds
to the regime $\varpi\gg m$.
\bibitem{Notedel} We shall assume $\delta>0$ without loss of generality.
\bibitem{Noteintg} Here we have used the relation $1=e^{-2i\pi n}$,
where the resonance parameter $n$ is an integer.
\bibitem{NoteGam} Here we have used equation 6.1.23 of \cite{Abram}
to infer the facts that
$\Gamma(2i\delta)/\Gamma(-2i\delta)=e^{i\phi_1}$ and
$\Gamma(1/2-i\delta-\kappa)/\Gamma(1/2+i\delta-\kappa)=e^{i\phi_2}$,
where $\{\phi_1,\phi_2\}\in\mathbb{R}$. These relations imply that
${\cal R}\in\mathbb{R}$.
\bibitem{Notemmu} Note that the large-mass $M\mu\gg1$ regime (\ref{Eq49}) corresponds to the asymptotic $m\gg1$
regime. This also implies the strong inequalities $\delta\gg1$ and
$k\gg1$ [see Eqs. (\ref{Eq61}) and (\ref{Eq63}) below].
\bibitem{Notese} Here we have used the approximated relation
$[1+e^{-2\pi(k-\delta)}]^{1/2i\delta}=1-ie^{-2\pi(k-\delta)}/2\delta+O[(e^{-2\pi(k-\delta)}/\delta)^2]$
\cite{Abram} in the asymptotic $k-\delta\gg1$ regime [see
\cite{Notemmu} and Eq. (\ref{Eq61}) below].
\bibitem{Noteklem} Here we have assumed the equality $l=m$ for the harmonic indices of the explosive
scalar field, a choice which minimizes the value of $K$ \cite{Hodpp}
and thus maximizes the value of the exponent $\delta-k$ of
(\ref{Eq52}) [see Eq. (\ref{Eq22})].
\bibitem{Notewn} It is worth emphasizing again that we assume the
relation $\omega_{\text{R}}\simeq \omega_{\text{c}}$ [see Eqs.
(\ref{Eq40}) and (\ref{Eq43})].
\bibitem{NoteqQd} Note that, in the eikonal large-mass $M\mu\gg1$ regime (which also
corresponds to the $m\gg1$ regime, see \cite{Notemmu}), one finds
the relation
$[{{e^{-2\pi(k-\delta)}}/{2\delta}}]'=[{{e^{-2\pi(k-\delta)}}/{2\delta}}]\times
[2\pi(\delta-k)'-\delta'/\delta]=[{{\pi
e^{-2\pi(k-\delta)}}/{\delta}}]\times (\delta-k)'[1+O(m^{-1})]$,
where here a prime denotes a derivative with respect to the
dimensionless variable $\gamma$. Thus, in the eikonal large-mass
$M\mu\gg1$ regime, the location of the maximum of $\delta-k$
corresponds [up to a small correction factor $1+O(m^{-1})]$ to the
location of the maximum of ${{e^{-2\pi(k-\delta)}}/{2\delta}}$ [see
Eq. (\ref{Eq52})].
\bibitem{Notedm} Note that the exponent of (\ref{Eq52}) is characterized by the
inequality $2\pi(\delta-k)<0$ in the entire range $s\in (0,1]$ of
the dimensionless black-hole rotation parameter (see also
\cite{Notese}).
\end{thebibliography}
\end{document}